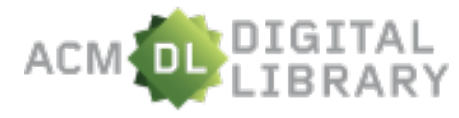
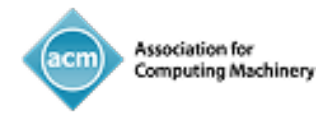
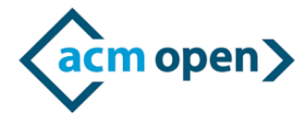

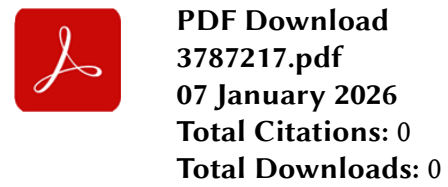

RESEARCH-ARTICLE

# S2Vec: Self-Supervised Geospatial Embeddings for the Built Environment

# S2Vec: Self-Supervised Geospatial Embeddings for the Built Environment

SHUSHMAN CHOUDHURY, Research and Machine Intelligence, Google Inc, Mountain View, United States
CHANDRAKUMARI SUVARNA, Research and Machine Intelligence, Google Inc, Mountain View, United States
IVEEL TSOGSUREN, Research and Machine Intelligence, Google Inc, Mountain View, United States
ABDUL RAHMAN KREIDIEH, Research and Machine Intelligence, Google Inc, Mountain View, United States
ELAD AHARONI, Research and Machine Intelligence, Google Inc, Mountain View, United States
CHUN-TA LU, Research and Machine Intelligence, Google Inc, Mountain View, United States
NEHA ARORA, Research and Machine Intelligence, Google Inc, Mountain View, United States

Scalable general-purpose representations of the built environment are crucial for geospatial artificial intelligence applications. This paper introduces S2Vec, a novel self-supervised framework for learning such geospatial embeddings. S2Vec uses the S2 Geometry library to partition large areas into discrete S2 cells, rasterizes built environment feature vectors within cells as images, and applies masked autoencoding on these rasterized images to encode the feature vectors. This approach yields task-agnostic embeddings that capture local feature characteristics and broader spatial relationships. We evaluate S2Vec on several large-scale geospatial prediction tasks, both random train/test splits (interpolation) and zero-shot geographic adaptation (extrapolation). Our experiments show S2Vec's competitive performance against several baselines on socioeconomic tasks, especially the geographic adaptation variant, with room for improvement on environmental tasks. We also explore combining S2Vec embeddings with image-based embeddings downstream, showing that such multimodal fusion can often improve performance. Our findings highlight how S2Vec can learn effective general-purpose geospatial representations of the built environment features it is provided, and how it can complement other data modalities in geospatial artificial intelligence.

## 1 Introduction

We present a highly scalable and modular approach for developing general-purpose representations of the built environment, i.e., human-made structures and surroundings that enable socioeconomic activity. Such

Authors' Contact Information: Shushman Choudhury, Research and Machine Intelligence, Google Inc, Mountain View, California, United States; e-mail: shushmac@google.com; Chandrakumari Suvarna, Research and Machine Intelligence, Google Inc, Mountain View, California, United States; e-mail: chandrasuvarna@google.com; Iveel Tsogsuren, Research and Machine Intelligence, Google Inc, Mountain View, California, United States; e-mail: iveel@google.com; Abdul Rahman Kreidieh, Research and Machine Intelligence, Google Inc, Mountain View, California, United States; e-mail: aboudy@google.com; Elad Aharoni, Research and Machine Intelligence, Google Inc, Mountain View, California, United States; e-mail: eaharoni@google.com; Chun-Ta Lu, Research and Machine Intelligence, Google Inc, Mountain View, California, United States; e-mail: chunta@google.com; Neha Arora, Research and Machine Intelligence, Google Inc, Mountain View, California, United States; e-mail: nehaarora@google.com.

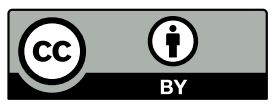

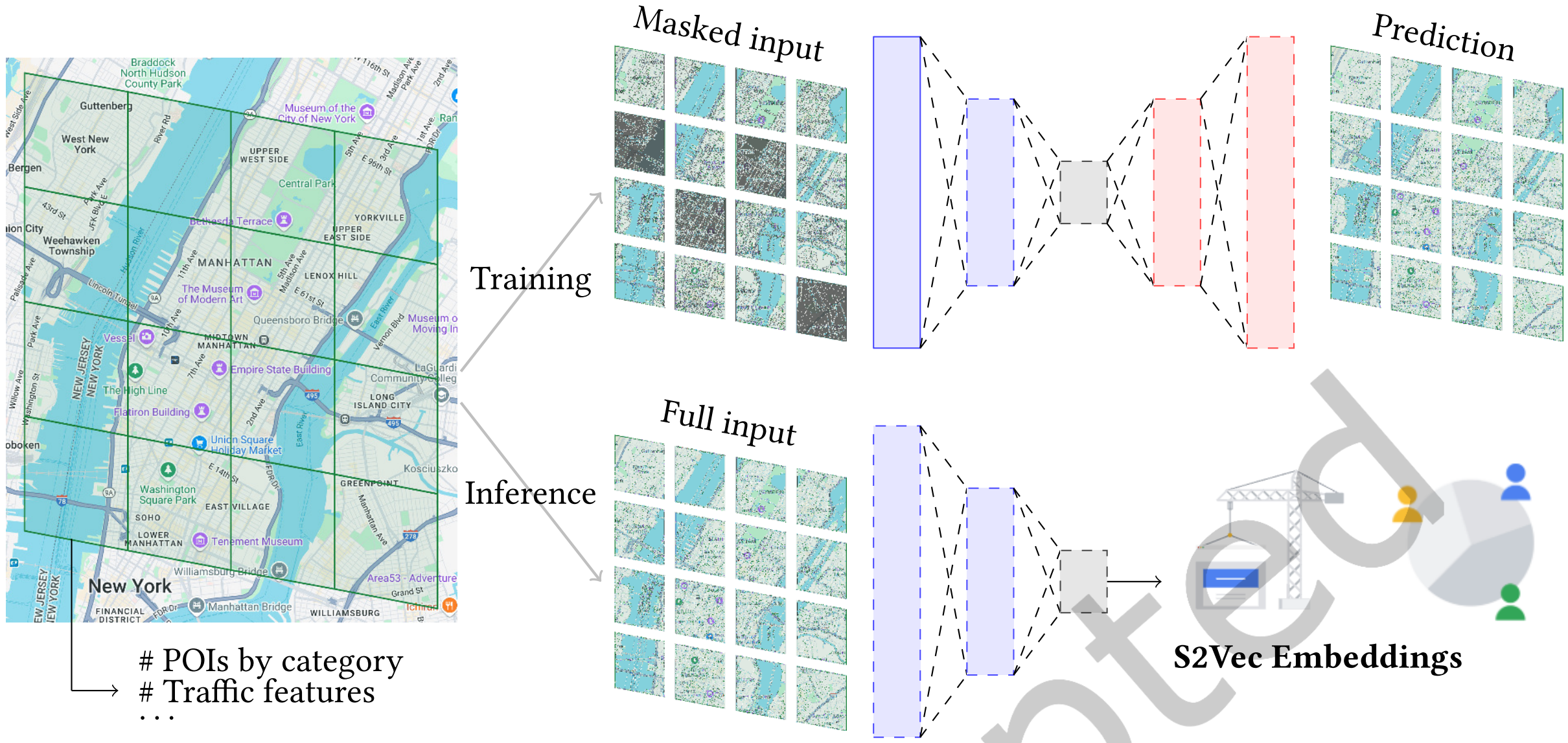

Fig. 1. For the S2Vec framework, we divide a large geographic area into fine-grained cells using the S2 library and construct a feature vector of built environment attributes for each fine-grained cell. We then rasterize the feature vectors as patches in an 'image' of a larger cell, and create a dataset of such images. Finally, we run masked autoencoding, a self-supervised learning approach on this image dataset to generate embeddings for the patches. The S2Vec embedding for any given location is thus the patch embedding of the fine-grained cell containing the location. We then evaluate it on predicting downstream socio-economic metrics such as housing prices and population density.

representations, or embeddings, can be adapted through downstream machine learning models for many tasks in geospatial artificial intelligence (or GeoAI) [22, 28]. Socioeconomic applications in urban settings require an understanding of the built environment that is not only fine-grained and precise but also task-agnostic and generalizable across diverse use cases and potentially sparse task-specific datasets [2, 19].

Learning spatial representations and encoding geographic locations is an active area of work within GeoAI, with many recent approaches and even libraries to standardize them [23, 40]. Broadly, these works have focused on the following aspects: encoding a point or a spatial object through analytical functions [24, 25], using aerial and street-level imagery at a location for learning image-based representations [5, 18], and encoding various task-relevant properties at given geographic coordinates [3, 29, 39]. Many of these approaches have proved effective on a range of geospatial classification and regression tasks.

Our key idea is twofold: first, we represent a geographic location through built environment features in a grid cell centred at that location; second, we learn embeddings that can capture the feature counts at a location and how they change over a larger area. For the first idea, we use the S2 geometry library [14], an optimized hierarchical geospatial index, to efficiently gather built environment feature vectors at cells of a desired resolution in any large area of interest (e.g., the United States). For the second idea, we rasterize the cell feature vectors as patches in a larger image, and use Masked Autoencoding [16] over the image dataset to learn and generate embeddings for every patch cell in the area of interest. Our approach, which we call S2Vec (Figure 1), is modular and customizable to any area of interest and any desired cell resolution.

We evaluate S2Vec by using its embeddings in a two-layer feed-forward neural network on multiple downstream geospatial regression benchmarks. Three of them are socioeconomic, i.e., California housing prices, US-wide population density and US-wide median income, while another two are environmental, i.e., US-wide tree cover and elevation. As baselines, we use three sophisticated image-based approaches, SATCLIP [18], GEOCLIP [5] and RS-MaMMUT, referring to the best performing of the MaMMUT models introduced in [4], to test how embeddings from built environment feature vectors compare with those from imagery. For the California housing task, we also compare with Hex2vec [39] and GeoVex [9] embeddings, both of which encode built environment features in different ways. Besides comparing with other embeddings, we explore various strategies to combine S2Vec with SATCLIP, GEOCLIP and RS-MaMMUT in a multimodal embedding in the downstream model. The built environment is inherently multimodal, and we want to investigate whether and to what extent these two modes can complement each other.

From our experiments on the three socioeconomic tasks, we find that S2Vec alone is competitive with the image embedding baselines on all benchmark tasks. Moreover, using multimodal embeddings typically performs better than or similar to either of the individual modes. When aggregating signals from multiple embeddings, we find that the choice of fusion strategy between the two modes does not significantly influence performance, though a particular one (project-then-add) is numerically the best fusion strategy across all settings. When we switch the task type to zero-shot geographic adaptation, i.e., where the held-out test set is chosen not at random but as a specific sub-region of the US, we find that S2Vec is typically the best individual mode and that fusing with the best image embedding creates the best overall approach. For the two environmental tasks, we find that S2Vec typically underperforms relative to the image-based embeddings; this finding highlights its current limitations due to the focus on the built environment and corresponding room for improvement. In summary, our paper makes the following contributions:

- An efficient and globally scalable scheme for representing locations through features of the built environment using the S2 Geometry geospatial index.
- S2Vec, a framework for learning task-agnostic embeddings of the built environment features, using self-supervised learning through masked autoencoding.
- Large-scale evaluations that compare and combine S2Vec with image-based embeddings on a range of downstream benchmark geospatial prediction tasks.

The rest of this paper is organized as follows. Section 2 discusses the S2 Geometry library and related work in spatial embeddings, masked autoencoding, and multimodal GeoAI. In Section 3 we describe our S2Vec framework of learning self-supervised geospatial embeddings in detail. Section 4 briefly motivates and discusses our multimodal fusion techniques. We lay out our extensive evaluation setup in Section 5 and reflect on the key themes of the findings in Section 6. Finally, Section 7 concludes with a summary and ideas for future work.

## 2 Background and Related Work

### 2.1 S2 Geometry

The S2 Geometry Library, developed by Google, is a comprehensive geospatial computation framework. It partitions the Earth into a hierarchy of cells, known as *S2 cells*, while modeling the Earth as a three-dimensional sphere rather than a flat two-dimensional projection. This design helps preserve spatial locality and enables efficient geometric operations with a single unified coordinate system and geographic database. Each S2 cell has a unique identifier, allowing for rapid lookups and spatial queries. The hierarchical structure supports multi-resolution analysis and makes it ideal for scalable geospatial applications. In our work, the S2 Geometry Library is fundamental to partitioning a given large geographic area into discrete cells, letting us efficiently extract and arrange built environment features while maintaining the spatial relationships necessary for accurate representation learning.

## 2.2 Location and Spatial Embeddings

How to represent geographic space effectively in machine learning models is a foundational challenge across GeoAI domains such as remote sensing, urban intelligence, and ecology [27]. A rich body of work has explored this problem [23] and yielded standard benchmarks and frameworks [40]. Three conceptual classes of approaches are relevant here. First are those that create high-dimensional projections of 2D or 3D coordinates or geometric objects that are learning-friendly, i.e., that task-specific neural networks can use effectively. Example techniques for this encoding approach use multi-frequency sinusoidal functions [24], Fourier transformations [25, 36], double Fourier spheres [29], and spherical harmonics [35].

Second are the approaches that generate embeddings from imagery (either satellite or geo-tagged photographs) using unsupervised or self-supervised contrastive learning techniques, e.g., MOSAIKS [34], CSP [26], SATCLIP [18] and GEOCLIP [5]. All of these methods have been effectively used in a range of downstream inference tasks. Their specifics and relative strengths vary, but they all benefit from the significant geospatial information contained in images and from powerful techniques in computer vision.

Third, and most relevant for us, are those that use structured attributes from a map, e.g., OpenStreetMaps (OSM) [15] to derive general-purpose representations that capture the semantics of locations. GeoVectors [38] provides a large-scale open corpus of OSM entity embeddings, using neural location embeddings to model spatial relationships and a bag-of-word model on OSM tags for semantics. CityFM [3] updates those ideas using more recent techniques, combining text and visual encoding of tags and geometries respectively. Perhaps the most similar approach to ours at a high level are Hex2vec [39] and GeoVeX [9]. Both learn embeddings from OpenStreetMap (OSM) feature vectors using Uber's H3 spatial index. We explicitly contrast S2Vec to them in Section 3.5, once we have described our methodology in detail.

## 2.3 Masked Autoencoding for Representations

Masked autoencoding (MAE) has emerged as a powerful paradigm for self-supervised learning, particularly in visual domains. The pioneering work of He *et al.* [16] showed how deliberately hiding portions of an image at random and training a network to recover the missing content can yield robust and scalable representations. MAE transforms the learning process into a puzzle-solving task by requiring the model to infer the hidden structure from contextual cues.

The Context Autoencoder [7] refines the core MAE idea by emphasizing the role of surrounding information and improving the model's ability to grasp the underlying semantics of the data. Recognizing that real-world data often spans multiple resolutions, Scale-MAE [32] introduces scale-awareness to capture features at varying levels of detail–a critical capability for applications involving geospatial data. In parallel, the Mixed Autoencoder framework [6] integrates diverse data types into a single reconstruction task, thus enabling more versatile and resilient feature representation. Multimodal Masked Autoencoders [12] jointly learn from different modalities, effectively aligning and transferring knowledge across disparate data sources.

Collectively, these and many other MAE innovations not only enhance our understanding of visual content but also inspire new approaches for capturing complex, multiscale, and multimodal information. We harness these principles in S2Vec to learn expressive geospatial representations of the built environment.

## 2.4 Multimodality in Geospatial AI

The built environment is inherently multimodal. To advance Geospatial AI, diverse data modalities have become a driving force, as they allow for rich, multiscale and nuanced representations of urban areas and human activity within them. Broad analyses on foundation models in GeoAI [22, 28, 45, 46] discuss both the promising opportunities and inherent challenges of this integrative approach. For example, combining satellite imagery, street-level views, and other helpful geospatial data can reveal spatial patterns that remain hidden with a single

**Algorithm 1** How S2Vec learns geospatial embeddings of the built environment

**Require:** Geographic Area $\mathcal{A}$, s2 patch level $l$, image level $l' < l$, embedding dimensionality $n$
**Ensure:** Embeddings $\phi(s_l)$ for all $s_l \in S_l$ (i.e., all level-$l$ cells covering $\mathcal{A}$)
1: Generate and define S2 cells sets $S_l, S_{l'}$ from $A$
2: **for** each $s_l \in S_l$ **do**
3: Generate feature vector $\Theta(s_l)$ ▷ Extract built environment features in the cell
4: **end for**
5: Create rasterized image dataset $D$, where $D(s_{l'}) = [\Theta(s_l^1), \Theta(s_l^2), \ldots]$ and $s_l^i \in \{\text{Children}(s_{l'})\}$
6: Create Masked Auto-Encoding model $M$ with patch encoder dimensionality $n$
7: Train model $M$ on image dataset $D$ ▷ Randomly masked patches are $\Theta(s_l)$ for some $s_l \in S_l$.
8: **for** each $s_l \in S_l$ **do**
9: Call patch encoding layer of trained M on $\Theta(s_l)$ to get $\phi(s_l)$
10: **end for**

modality [30]. The multimodal fusion enriches the learned representations and includes details that are otherwise overlooked, yielding predictive insights into critical socioeconomic indicators such as income levels, overcrowding, and environmental deprivation [37].

Multimodal data also benefits from creative machine learning techniques. For instance, the SeMAnD framework [33] uses self-supervised anomaly detection across different geospatial datasets, highlighting the strength of multimodal data in uncovering irregular urban patterns that signal significant changes. Recent research uses heterogeneous graph-based embeddings [47] to model urban regions as interconnected nodes, effectively capturing the complex interplay among various urban elements. Additionally, trajectory-powered models [8, 20] highlight the importance of integrating mobility data with other geospatial signals to create a comprehensive picture of movement within cities.

## 3 Self-Supervised Geospatial Embeddings

### 3.1 Problem Definition

We start with a given large geographic landmass of interest $\mathcal{A}$, e.g., the continental United States of America. We are also given the level $l$ of the S2 cells into which to divide $\mathcal{A}$, e.g., in all our experiments we use level 12, for which each cell covers a surface area of approximately 5 $\text{km}^2$.

Let $\mathcal{S}_l$ be the set of all non-overlapping level-$l$ S2 cells that covers the area. Then the ultimate goal of S2Vec is to learn a mapping $\Phi : \mathcal{S}_l \rightarrow \mathbb{R}^n$, where $n$ is the user-specified dimensionality of the embedding. Given any pair of GPS coordinates, S2Vec efficiently looks up the unique level-$l$ S2 cell that contains it, e.g., $s_l$, and returns $\Phi(s_l)$, the corresponding embedding vector.

S2Vec is designed to be a modular and globally scalable pipeline for encoding properties of geographic locations into fixed-dimensional embeddings. It does so by leveraging efficient hierarchical geospatial indexing and rasterization of built environment feature vectors, and self-supervised representation learning via masked autoencoding (Algorithm 1). The resulting task-agnostic embeddings can then be directly used or aligned with image-based representations in downstream task-specific machine learning models for various geospatial inference applications.

### 3.2 The Feature Vector

We start with the set of level-$l$ S2 cells (i.e., $\mathcal{S}_l$) into which the large landmass $\mathcal{A}$ has been efficiently partitioned by the S2 Geometry Library. To capture the spatial distribution of built environment features, we represent each S2 cell $s_l$ with a feature vector $\Theta(s_l)$. For the experiments in this paper, this vector is a histogram of counts that includes:

- **Place-of-Interest (PoI) Categories:** Counts of various categories of geographic entities within the cell, based on a map's ontology (e.g., number of shops, restaurants, gas stations, beauty services and so on). We use Google Maps' internal database for our experiments, but our taxonomy has considerable overlap with OpenStreetMaps too.
- **Road Network Features:** Counts of infrastructural elements such as roads, traffic lights, and other relevant markers.

We use raw feature counts for this initial work rather than more complex representations (e.g., normalized densities, TF-IDF weighting, or pre-trained category embeddings) in order to isolate the performance of S2Vec's learning approach. By using simple counts, we can *better attribute the model's success* to the power of the masked autoencoding on the rasterized spatial data rather than to elaborate feature engineering. An attention-based model like the Vision Transformer is capable of implicitly learning the relative importance of different features during its reconstruction task. For instance, it can determine from the data that a rare PoI tag is more informative than a common one, achieving a similar outcome to an explicit weighting scheme like TF-IDF [42].

A known challenge with any geospatial database (even one as rich as the Google Maps network that we use) is the potential for out-of-date information or coverage bias, where some areas are mapped more thoroughly than others. While we acknowledge that S2Vec's performance depends on the quality of its data, the underlying problem of map completeness and freshness is outside the scope of this work. That said, our methodology is inherently robust to some of these issues. For out-of-date PoIs, a straightforward approach is periodic retraining; the S2Vec data processing pipeline is highly scalable and parallelizable, making it easy to regenerate feature vectors and embeddings as the underlying map data evolves. More importantly, the Masked Autoencoding (MAE) learning paradigm is natively robust to sparsity within any given child cell. By learning to reconstruct a patch from its neighbors, the model can *infer the characteristics of a sparsely-mapped cell* from the context of its potentially better-mapped surroundings.

Our framework can also differentiate between true sparsity, e.g., a rural area with few amenities, and domain-specific sparsity. The fact that a PoI category like nightlife is naturally less common outside of urban cores can be a valuable spatial signal that S2Vec is designed to capture. The model learns the contextual significance of both the presence and absence of features. Finally, the S2Vec framework is agnostic to the specific composition of the feature vector. We chose a simple and effective infrastructure count histogram for our experiments, but this vector could be arbitrarily extended to other numerical features.

## 3.3 Rasterizing Feature Vector Images

Once we have a feature vector for every level-$l$ base cell, the next step is to arrange groups of adjacent $s_l$ cells as patches in images. The set of all such images will be a *disjoint partition* of $\mathcal{S}_l$, i.e., no two images will have any overlapping $s_l$ patches, and the union of all images will use every cell in $\mathcal{S}_l$. We leverage the hierarchical indexing of the S2 Geometry Library to do this effectively. We pick a lower level of S2 cells, i.e., $l' < l$, and divide $\mathcal{A}$ into $\mathcal{S}_{l'}$, the corresponding set of level-$l'$ cells. By construction, each level-$l'$ comprises exactly $2^{l-l'} \times 2^{l-l'}$ level-$l$ so-called child cells. We map these child $s_l$ cells to their corresponding 2D location within $s_{l'}$ and rasterize the $\Phi(s_l)$ feature vectors (i.e., treating each element as a 'pixel' of an image patch unrolled into a 1D vector). As a result, every parent $s_{l'}$ cell now corresponds to a rasterized image of the feature vectors of its child $s_l$ cells, i.e., $D(s_{l'}) = [\Theta(s_l^1), \Theta(s_l^2), \ldots]$.

The fixed row-by-row ordering preserves the spatial continuity and local context of built environment features and ensures that the model's subsequently learned positional embeddings correspond to geographic relationships between the cells. Our model architecture is technically agnostic to the initial patch order. But if we gave it a scrambled ordering of patches, the learned positional embeddings would have to model that complex permutation. Given that the semantic meaning of the whole is derived from the correct spatial arrangement of its parts (as in images, where scrambling the patches would alter the image meaning or make it meaningless), a geographically consistent ordering is the right approach.

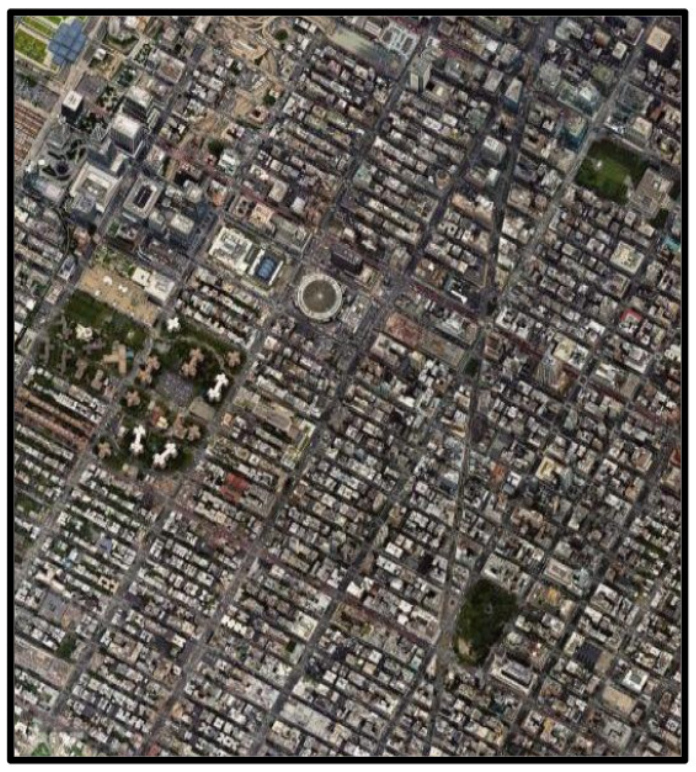
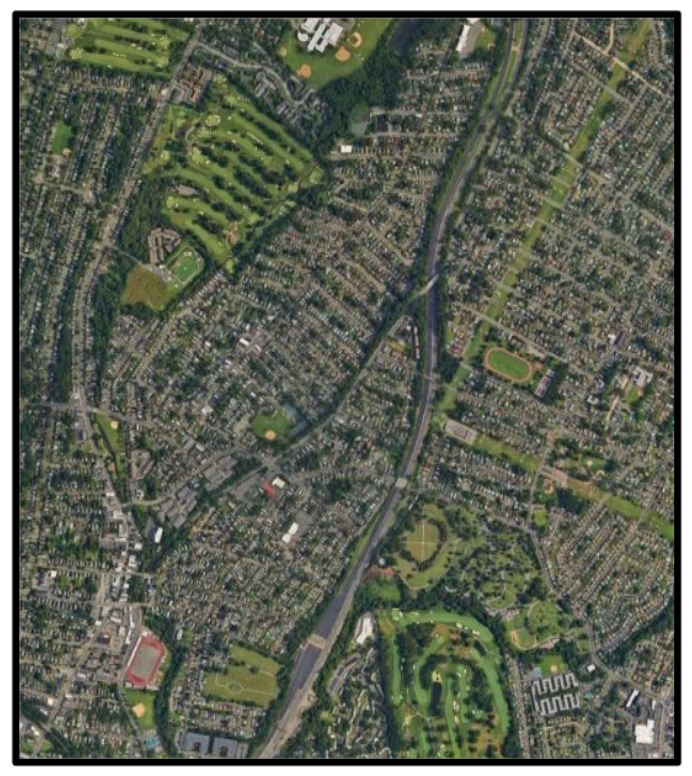
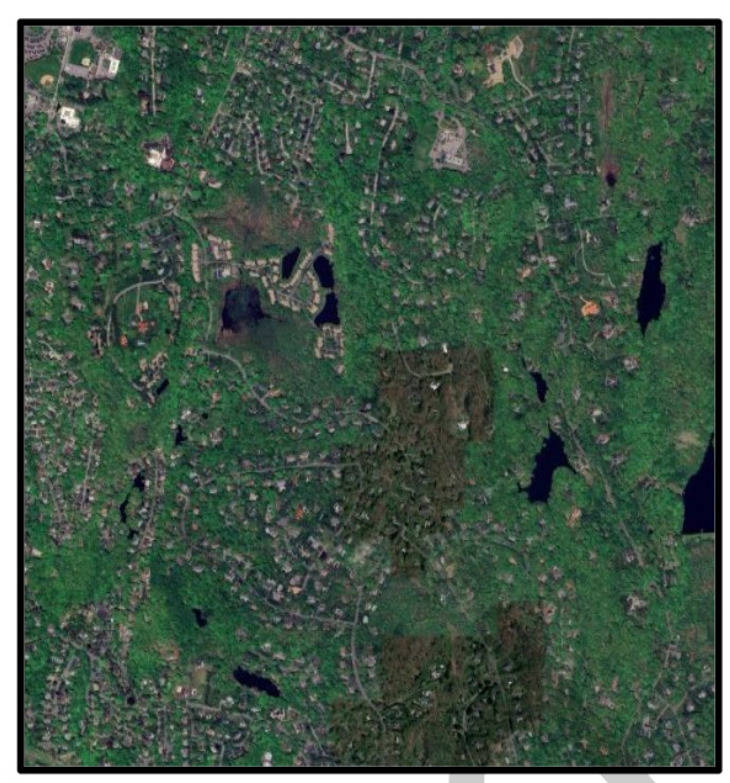

Fig. 2. Three distinct kinds of built environment areas that help illustrate how S2Vec and image-based representations could complement each other. The left (downtown) and middle (suburb) would have some overlap in feature vector space, as would the middle and right (rural). But overhead imagery could help distinguish different kinds of suburbs (for instance) depending on the spatial arrangement of the built environment features. The images are taken from Google Maps' satellite view.

## 3.4 Learning Embeddings with Masked Autoencoding

After constructing the rasterized feature vector images, we apply masked autoencoding (MAE) to learn embeddings. This grid-based approach explicitly preserves the spatial relationships between adjacent geographic cells, allowing MAE to leverage rich spatial contexts [11]. By masking random patches and reconstructing them, the model transcends simple feature counting to learn the fundamental semantics of the built environment—such as how commercial, residential, and transport features co-locate to form patterns like downtowns or suburbs. S2Vec also uses self-attention to model long-range dependencies across the entire parent S2 cell (covering 1300 $\text{km}^2$ in our experiments). This enables it to capture complex patterns extending beyond local adjacencies, such as the gradual transition from an urban core to the periphery. By effectively capturing the grammar of the built environment, the resulting embeddings act as powerful, generalizable representations for diverse downstream tasks.

Our implementation follows the original MAE paper [16], which is based on the influential Vision Transformer [10]. Below are the broad steps of the MAE in S2Vec (we adapt the language of Section 3 of the original MAE paper where relevant):

- **Random Masking:** Each image has $2^{l-l'} \times 2^{l-l'}$ patches, each corresponding to a feature vector $\Theta(s_l)$ of some child $s_l$. We sample a subset of patches uniformly at random without replacement and mask (remove) the other ones. The masking of a given image changes each time the image is observed across epochs, which is also a form of data augmentation.
- **Encoding**: This step uses a Vision Transformer on the unmasked patches. The patches are embedded with a standard linear projection and learned positional embeddings and are processed by a set of Transformer blocks. No mask tokens are required as the masked patches are removed. We will also reuse the patch embedding at the end of the process.

- **Decoding:** The decoder is another set of Transformer blocks. It is given both encoded unmasked patches as well as mask tokens (learned vectors that identify which patches are missing and need to be predicted). A key benefit of MAE is the independence of the decoder relative to the encoder, because the former is only used during the self-supervised stage and not for generating the image representations (and in our case, for the patch embeddings). This independence allows it to be typically small and makes the full training much more compute-efficient.
- **Self-Supervision:** The core learning signal comes from a self-supervised reconstruction objective. The model minimizes the difference between the original patches and its reconstructions of the masked ones. This objective guides the network to capture essential spatial patterns and makes it robust to incomplete or noisy inputs. The resulting embeddings will capture both explicit feature counts and implicit spatial correlations, making them highly transferable for many downstream tasks.
- **Generating Patch Embeddings:** After the MAE training is over, we take all $s_l$ cells, lookup their $\Theta(s_l)$ feature vectors, and run the trained patch encoder on each $\Theta(s_l)$ to get the corresponding $\Phi(s_l)$ embedding. In contrast, the original MAE model, when used on image recognition, encodes the entire image to get its latent representation. We do not care about image-level ($l'$) representations, only patch-level ($l$) ones. But if a downstream application would benefit from embeddings for coarser-grained image-level cells, i.e., $\Phi(s_{l'})$, then S2Vec would be suitable for those as well.

### 3.5 Conceptual Novelty

S2Vec diverges from prior grid-based embedding methods, most notably Hex2vec [39] and GeoVeX [9], in its learning objective, architectural backbone, and inference efficiency.

Hex2vec uses a contrastive approach adapted from the Skip-gram model, where the objective is to discriminate between positive pairs (immediate neighbors) and negative samples. As a result, its context is explicitly local, limited to the first ring of H3 neighbors. In contrast S2Vec learns to reconstruct the high-dimensional feature vectors of masked regions instead of relying on fixed neighbor sets. By rasterizing fine-grained cells into large parent images, S2Vec models long-range, multi-scale dependencies across the area, capturing macro-level patterns that local window methods like Hex2vec would miss.

While GeoVeX also uses autoencoding, it relies on a Hexagonal Convolutional Autoencoder (HCAE) with fixed-weight kernels. S2Vec leverages a Vision Transformer, using self-attention to dynamically weigh the importance of contextual patches based on the data itself. This architectural difference leads to differences in efficiency. GeoVeX requires redundant computation: it must encode a large neighborhood (7 rings) to produce an embedding for only the central cell. S2Vec generates embeddings for all child cells within a parent image in a single forward pass. Finally, while GeoVeX requires a specialized Zero-Inflated Poisson (ZIP) loss to handle sparse data, S2Vec demonstrates that standard normalization combined with the robust MAE paradigm effectively handles sparsity without domain-specific loss functions [44].

## 4 Fusing with Image Embeddings

The data modality captured in S2Vec is essentially a geospatial knowledge database, which is used to generate the feature vectors of counts for each S2 cell. This is one of several modalities that are useful and relevant for GeoAI applications; others include plain text, street view and remote sensing images, mobility data, and vector representations ([22]; Section 4). Of these other modes, imagery in particular has the potential to align with and complement S2Vec (illustrated in Figure 2).

To empirically assess the benefits of such multimodal representations, we explore fusing the S2Vec embeddings with separately trained image-based embeddings in the downstream task-specific models (we use three different image-based embeddings as mentioned in Section 5.3). We try three different commonly-used strategies for the two sets of embeddings:

(1) **Concatenation:** We concatenate the embedding vectors to form a joint representation whose length is the sum of the two embedding dimensionalities.
(2) **Weighted Addition:** We apply learnable weights to each embedding before element-wise addition; this only works if the two sets of embeddings have the same dimensionality.
(3) **Projection and Addition:** We separately project each embedding with a hidden layer to the same dimensionality and then add them element-wise.

These techniques will allow us to examine how different fusion methods impact performance on downstream tasks, and whether and to what extent the built environment and image-based representations can complement each other.

## 5 Evaluation

### 5.1 S2Vec Implementation

We select the level of an image S2 cell to be 8 and that of a patch cell to be 12, a choice that balances geographic detail, contextual scope, and computational cost. A patch-level cell of 12, covering approximately 5 $\text{km}^2$, offers a resolution fine-grained enough to capture meaningful variations in local socioeconomic characteristics. The image-level cell of 8, covering approximately 1300 $\text{km}^2$ provides a broad contextual view—often encompassing a large portion of a metropolitan area—that is suitable for learning larger-scale spatial dependencies between the patches. This configuration results in a $2^{12-8} \times 2^{12-8} = 16 \times 16$ grid of patches per image, which is also a standard configuration for typical images.

We implement a distributed parallel processing pipeline to generate the feature vectors and images. This pipeline begins by querying Google Maps Road Network data (e.g., lanes, Points Of Interest) to retrieve the environment features for each patch level S2 cell. For each patch level cell, identified by its unique key, we construct a histogram feature vector of size 116, which includes counts of both specific point-of-interest categories and traffic infrastructure features in the cell (see Section A for some visualizations). To enable normalization in subsequent steps, we also compute the column-wise mean and variance across all the generated patch level feature vectors.

Subsequently, we use the S2 Geometry Library to rasterize the chosen image level S2 cells. This rasterization involves ordering the patch level cells within each image level cell row by row, starting from the top-left corner and proceeding to the bottom-right. To obtain the final unrolled feature vector for an image cell, we concatenate the individual histogram feature vectors of all the ordered patch cells into a single vector. Using this pipeline, we generate an image dataset for the entire US, comprising around 12000 images. Our globally scalable pipeline uses Google's proprietary internal Apache Beam-based infrastructure with C++ for the raw features and Python for the rasterization.

**Pre-training Details:** Prior to pre-training, we globally normalize all feature vectors with feature-wise means and variances. S2Vec uses the core masked autoencoding architecture from the original paper, but with specific hyperparameter settings tailored for our task. We adopt the AdamW optimizer [21] with a cosine decay learning rate schedule to maintain consistency with the original Masked Autoencoding paper.

We run our hyperparameter tuning on Vizier, a robust internal service at Google for black-box optimization [13]. The tuning process is validated on a randomly held-out 20% of the training data. Vizier's default algorithm uses Batched Gaussian Process Bandits, a Matérn kernel with automatic relevance determination, and the expected improvement acquisition function. After identifying the optimal set of hyperparameters, we re-train the final model on the full dataset.

Table 1. S2Vec Pre-Training Hyper-parameters

| Hyperparameter | Value | Tuning Range |
|---|---|---|
| **MAE Architecture** | | |
| Attention Heads | 8 | (2, 4, 8, 12) |
| Encoder Layers | 6 | (2, 4, 6, 8) |
| Decoder Layers | 2 | (1, 2, 3, 4) |
| Encoder Dimension | 256 | (128, 256, 512) |
| Decoder Dimension | 128 | (64, 128, 256) |
| **Optimizer** | | |
| Name | AdamW | Fixed |
| Weight Decay | 1e-3 | (1e-4, 1e-3, 1e-2) |
| Clipnorm | 1.0 | (0.5, 1.0, 2.0) |
| Initial Learning Rate | 5e-4 | (5e-5, 1e-4, 5e-4, 1e-3) |
| Alpha ($\alpha$) | 0.1 | (0.05, 0.1, 0.2) |
| **Training** | | |
| Batch Size | 256 | (64, 128, 256, 512) |
| Shuffle Buffer Size | 1000 | Fixed |
| Number of Epochs | 50 | Fixed |
| Dropout Rate | 0.2 | (0.1, 0.2, 0.3) |
| Mask Ratio | 0.5 | (0.25, 0.5, 0.75) |

Throughout all experiments, we fix the embedding dimensionality at 256. We chose this standard size to provide sufficient representational capacity for downstream models while reducing the risk of overfitting. This dimension also aligns with that of baseline methods, ensuring a fair comparison. An example of clustered embeddings is provided in Appendix B. We use Tensorflow 2.0 and Keras with Python for all the model code. The pre-training hyperparameter tuning and final runs are all done on 8 Nvidia V100 GPUs, with distributed training using Tensorflow's mirrored data parallelism strategy [1]. Table 1 summarizes the final hyperparameters and the ranges explored during tuning.

The S2Vec implementation used for our experiments is proprietary to Google Research and cannot be shared publicly. An open-source version has been made available in the SRAI library[1] (no authors were involved in this implementation).

## 5.2 Downstream Tasks and Models

We ran experiments on five diverse and representative geospatial prediction tasks. Three of them were socioeconomic attributes: California housing prices [31], US-wide population density [34], US-wide median income[17]. The other two were environmental attributes: tree cover and elevation, both US-wide [2].

These datasets differ in geographic resolution—while the first two provide values at specific coordinates, the remaining three aggregate data at the ZIP code level. To ensure consistency across datasets, we lined up all data points to level 12 S2 cells. This gave us approximately 6700 data points for the California housing prices dataset, 47000 data points for US-wide population density, 1.7 million data points for US median income, and 35000 datapoints for each of the US-wide environmental tasks. In cases where multiple data points fall within the same S2 cell, we assign the median label for that cell.

[1]https://github.com/kraina-ai/srai/tree/main/srai/embedders/s2vec

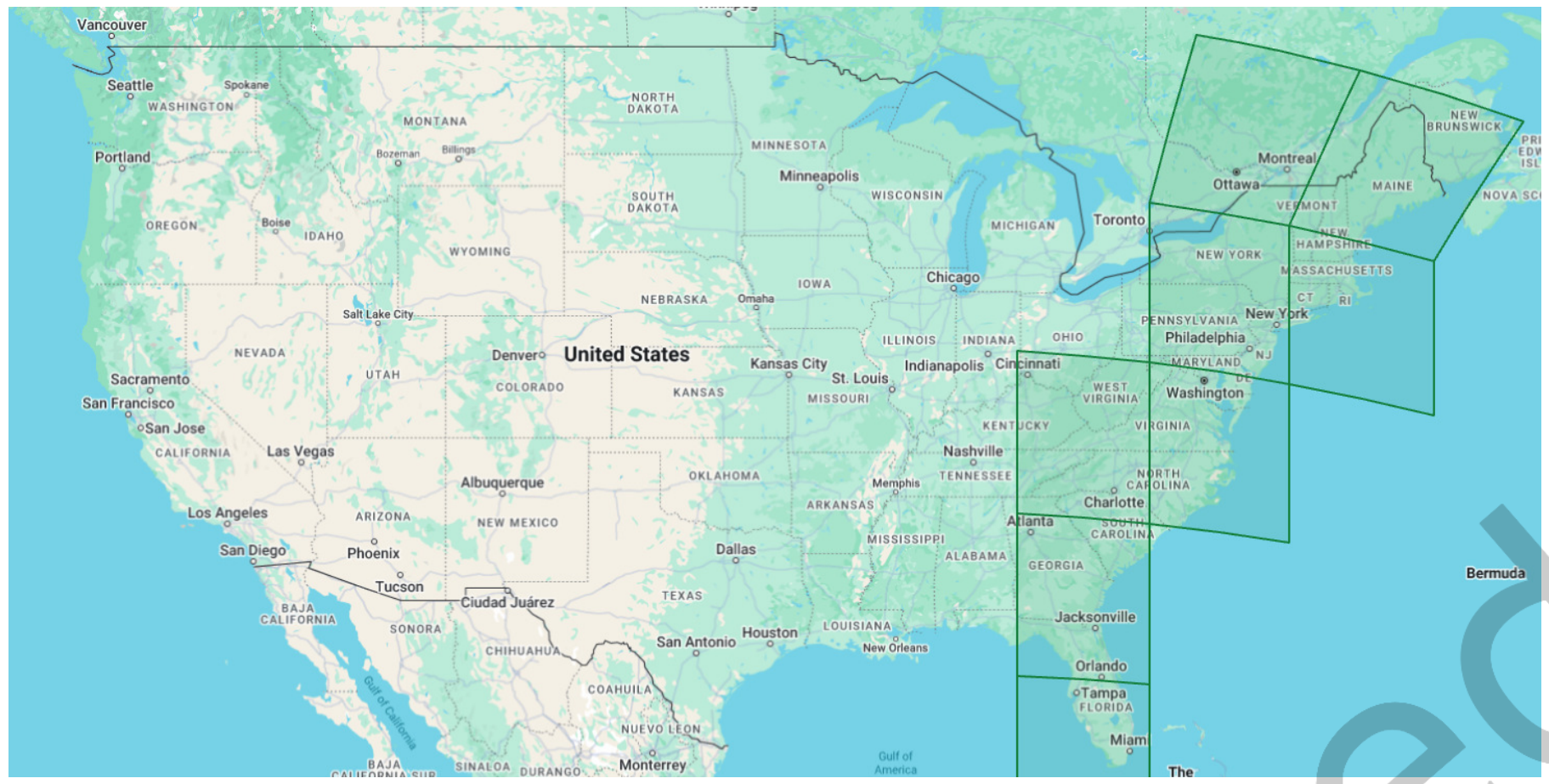

Fig. 3. The eastern portion of the US that we hold out in geographic adaptation experiments is covered by the green cells. This area is used as the final test set to evaluate the downstream models after training and validation on the remaining area.

For downstream predictive modeling, we use a two-layer multilayer perceptron on top of the learned S2Vec embeddings to predict the regression targets (min-max scaled between 0 and 1). We trained each downstream model using the mean squared error loss function and AdamW optimizer, and tuned the hyperparameters by sweeping over a predefined set of values for the learning rate, number of hidden units, and dropout rate on an independent validation set. We then use the best-performing set of hyperparameters to train the model on the training set, with early stopping based on the validation set loss to prevent overfitting. Finally, we evaluate the model on the held-out test set. This entire end-to-end process is repeated with multiple random seeds for more robust results. The scheme described here is standard for evaluating general-purpose embeddings and is similar to that used for evaluating SATCLIP [18]. The downstream experiments are conducted with a Python Colab notebook on a single V100 GPU.

### 5.3 Methods Compared

Our experiments compare several different methods and variations; we discuss them here in stages.

**Unimodal Embeddings:** First, we compare S2Vec individually with other image-based embeddings:

- SATCLIP: A contrastive vision-language model tailored for remote sensing, aligning satellite imagery with textual descriptions [18]. We use the ResNet18 backbone with L=40 legendre polynomials; the embeddings are 256 in length.
- GEOCLIP: A CLIP-inspired geo-localization model that aligns remote-sensing imagery with encoded location [5]. We use the default open source setting that produces embeddings that are 512 in length.
- RS-MaMMUT: A pre-trained RGB remote sensing vision-language model based on the MaMMUT architecture [4]. The embeddings are 1152 in length.

For the California housing task, we also compare S2Vec against the two closely related embedding approaches that use place-of-interest category tag counts from OpenStreetMaps with Uber's H3 index: Hex2vec [39], which does skip-gram modeling with negative sampling, and GeoVex [9], which uses a Hexagonal Convolutional

Autoencoder. We use SRAI's implementation[2] for both of these approaches; neither of their implementations are optimized for very large areas, so we only use them for the California-scale task.

**Adding Location Signal:** Neither S2Vec nor RS-MaMMUT use the location (i.e., the numerical coordinates or some derived form of them) of the S2 cells or the image in their pre-training. SATCLIP and GEOCLIP, however, explicitly encode the coordinates of the images in pre-training. Socioeconomic metrics can vary considerably based only on location, e.g., the median income in a zipcode in one US state may be very different from a zipcode in another even if their overhead imagery and built environment features are similar. Thus, to enable a fairer comparison with SATCLIP and GEOCLIP on the two US-wide tasks, and more broadly to evaluate to what extent the location signal helps S2Vec and RS-MaMMUT we have variants that include a location encoding (named S2Vec-Loc and so on). We do this by generating the Space2Vec analytical encoding [24] for the centroid of the s2 cell of each datapoint and concatenating it with the respective embeddings in the downstream model. This encoding is just one simple way to add the location signal, and exploring this further or optimizing it for a downstream task is out of the scope of this work.

**Multimodal Fusion:** Finally, we combine S2Vec with each image-based embedding in turn to evaluate the multimodal fusion; they are named S2-SATC, S2-GEOC, and S2-RS for conciseness. As discussed in Section 4, we try up to three different strategies (named concat, wt-add, and proj-add for conciseness). The wt-add variant, i.e., addition with learnable weights, only works for S2-SATC as they have the same length. For S2-RS, we also have a variant with the location signal included.

As with the task datasets, we made sure everything was lined up with level 12 S2 cells. The SATCLIP and GEOCLIP models give us embeddings for specific points. We generated embeddings that matched the locations in our datasets and used those in our prediction tasks. If an S2 cell had multiple geographic coordinates, we used the average of their embeddings. For RS-MaMMUT embedding generation, we process 1020x1020 pixel image crops centered on each level 12 S2 cell. These patches, representing a 2m/pixel ground resolution, are extracted from high-resolution satellite imagery between March 2021 and March 2025.

### 5.4 Results

We use a similar downstream evaluation scheme as SATCLIP [18]. For a given dataset and a given embedding method, we define the downstream model using the embedding and the best set of hyperparameters for that method (as described in Section 5.2). Then, over 20 random train/validation/test splits of the dataset (using a 60/20/20 split), we train the downstream model on the training set, use early stopping based on the validation set, and evaluate on the held-out test set.

For the 2 US datasets, we run an additional set of experiments on so-called *zero-shot geographic adaptation*, or how well the embeddings generalize across space. Here the train/validation/test split is not random but spatial, i.e., we hold out an entire sub-geography and use the remaining area for training and validation. For our experiments, we holdout the northern and eastern parts of the USA as shown in Figure 3. For this task type, since we use the location coordinates to split the data, we do not evaluate the variants of S2Vec or RS-MaMMUT with the location encoding.

We report two evaluation metrics on the held-out test set: the $R^2$ metric and the Mean Absolute Error metric. The $R^2$ metric (typically between 0 and 1; higher is better) is a goodness-of-fit metric that measures how well the regression predictions approximate the real data points. The Mean Absolute Error (Mean Abs Err; lower better) is a ubiquitous error metric in machine learning; here we report the Mean Abs Err for the normalized regression target (between 0 and 1) to make it more interpretable. Both metrics are informative and useful; we vary which one we report across experiments.

[2]https://github.com/kraina-ai/srai/tree/main/srai/embedders

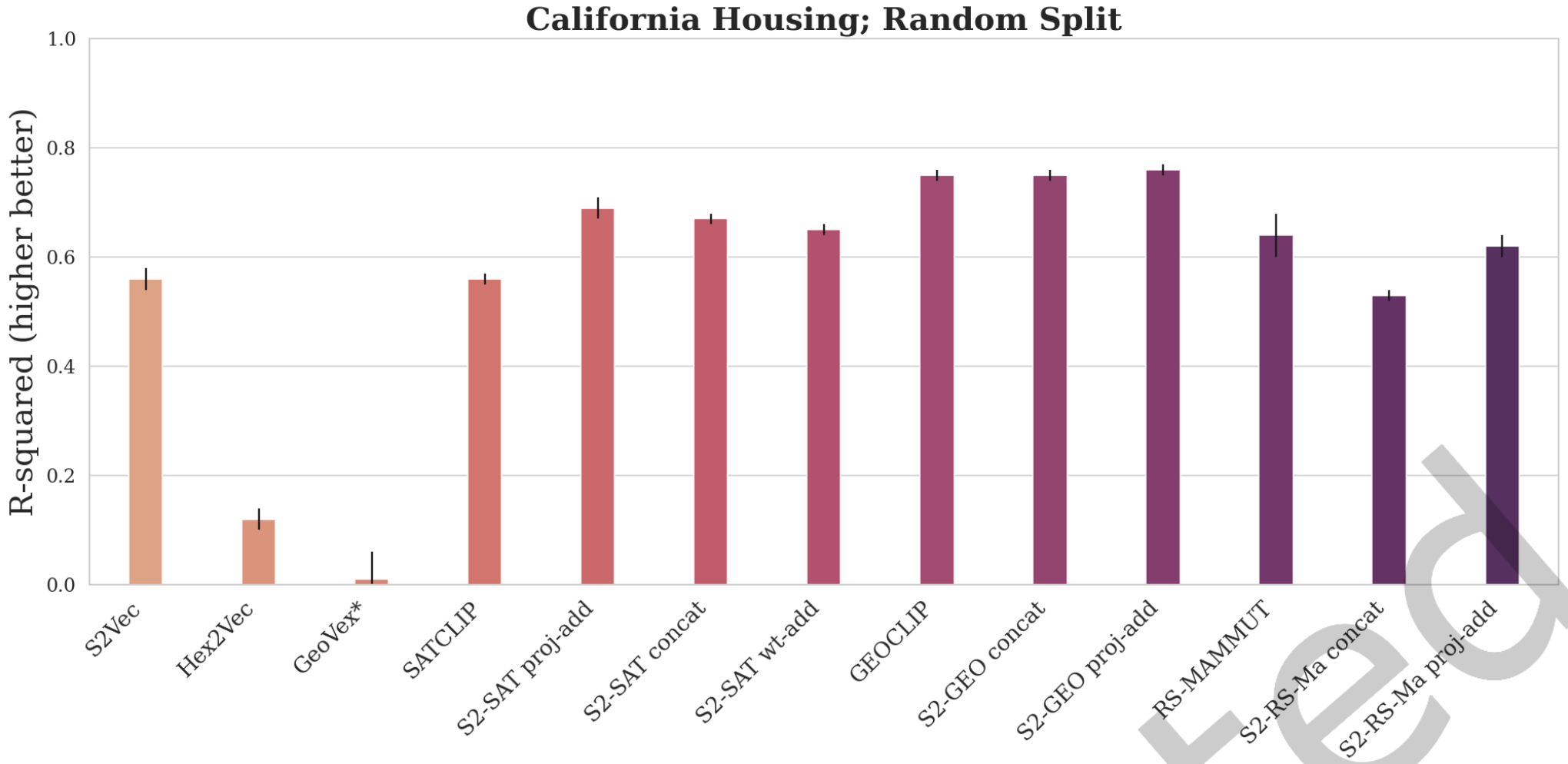

Fig. 4. On the California house price task, for individual modes S2Vec is comparable with SATCLIP, but RS-MaMMUT is better and GEOCLIP is the best. Fusing S2Vec with SATCLIP/GEOCLIP outperforms either mode, except with RS-MaMMUT where the fusion is likely overfitting on the relatively small dataset. GeoVex*: $R^2$ values are lower than $-1$ across many runs; we clip to 0 for visualization.

*5.4.1 Random Train/Val/Test Split.* The bar plot in Figure 4 shows how the various approaches compare in terms of the $R^2$ metric on the California housing price task. Among the unimodal embeddings, GEOCLIP performs the best, which is consistent with the benchmarking from the earlier SATCLIP paper; our metrics for GEOCLIP and SATCLIP for this task almost exactly match those in the SATCLIP paper([18]; Table 2). Then comes RS-MaMMUT, and then close behind are S2Vec and SATCLIP, both of which are comparable to each other. Both Hex2vec and GeoVex appear to perform significantly worse than any other approach. For GeoVex in particular, the $R^2$ metric can be negative and lower than $-1$, but we mark it as 0 in the plot. This finding could also be due to how they are implemented, and it is possible that extensive tuning of them on this task would improve performance (out of the scope of this work). As stated earlier, we do not compare against them for the remaining US-wide tasks.

Generally, methods combining S2Vec with other modalities (S2-SAT, S2-GEO) tend to outperform either of the individual modes. The exception to this is with RS-MaMMUT, where the multimodal alignment hurts performance. However, this is most likely due to overfitting on the relatively small dataset of 6600 datapoints and the higher embedding dimensionality of 1152 for RS-MaMMUT (compared with 256 for S2Vec/SATCLIP and 512 for GEOCLIP). Finally, the project-add approach (proj-add) is the best for multimodal alignment regardless of which two methods are used. This relative superiority of proj-add was repeated for the remaining tasks; for better readability, **subsequent results will only report proj-add among the fusion approaches**.

Next, Table 2 reports Mean Absolute Error of the various methods on the larger and more robust US-wide tasks. First, we discuss the two socioeconomic tasks: Population and Median Income. For each column, the method with the lowest Mean Absolute Error is boldfaced. On the Population task, RS-MaMMUT alone outperforms all other approaches, including the multimodal fusions, while S2Vec is second on individual modes. For each of SATCLIP and GEOCLIP, fusing with S2Vec through project-add improves on both of them individually (though not on S2Vec alone), while fusing S2Vec with RS-MaMMUT improves on the former but not the latter. Adding location signal for this task does not appear to improve any of the approaches to which it is added.

Table 2. Random train/validation/test split performance on the two US-wide datasets: population and median income. Over 20 independently initialized training runs, we report the mean and standard deviation of the Mean Absolute Error (lower better) over the held-out test set, min-max scaled between 0 and 1.

| Method | Population Mean Abs Err ↓ | Median Income Mean Abs Err ↓ | Tree Cover Mean Abs Err ↓ | Elevation Mean Abs Err ↓ |
|---|---|---|---|---|
| *Unimodal* | | | | |
| S2Vec | 0.065 ± 0.002 | 0.057 ± 5e-5 | 0.195 ± 6e-4 | 0.067 ± 5e-4 |
| SATCLIP | 0.075 ± 0.001 | 0.044 ± 1e-3 | 0.104 ± 7e-4 | 0.015 ± 3e-4 |
| GEOCLIP | 0.075 ± 0.005 | **0.033** ± **6e-5** | 0.091 ± 3e-4 | **0.015** ± **2e-4** |
| RS-MaMMUT | **0.057** ± **4e-4** | 0.061 ± 2e-3 | 0.092 ± 9e-4 | 0.045 ± 7e-3 |
| *With Location* | | | | |
| S2Vec-Loc | 0.065 ± 3e-4 | 0.054 ± 5e-5 | 0.185 ± 4e-4 | 0.058 ± 4e-4 |
| RS-MaMMUT-Loc | **0.057** ± **3e-4** | 0.058 ± 2e-3 | **0.09** ± **5e-4** | 0.044 ± 5e-3 |
| *Multimodal* | | | | |
| S2-SATC proj-add | 0.065 ± 0.003 | 0.045 ± 2e-3 | 0.113 ± 1e-3 | 0.018 ± 3e-4 |
| S2-GEOC proj-add | 0.066 ± 0.003 | 0.045 ± 3e-3 | 0.10 ± 9e-4 | 0.0194 ± 3e-4 |
| S2-RS proj-add | 0.058 ± 4e-4 | 0.056 ± 2e-4 | 0.102 ± 7e-4 | 0.048 ± 1e-3 |
| S2-RS proj-add-loc | **0.057** ± **3e-4** | 0.055 ± 1e-4 | 0.101 ± 5e-4 | 0.048 ± 1e-3 |

On median income, however, GEOCLIP performs the best overall. Fusing S2Vec with SATCLIP, GEOCLIP, and RS-MaMMUT improves on S2Vec in all cases and on RS-MaMMUT, though not on SATCLIP or GEOCLIP. Unlike in the previous dataset, adding the location signal here improves on all three approaches to which it is added. This difference supports the intuition that the median income task depends strongly on the underlying geographic location regardless of the built environment or the overhead imagery. This location dependence also helps explain the relatively strong performance of GEOCLIP and SATCLIP, which encode location explicitly during pre-training (rather than just combining them downstream in a lightweight manner).

Next, we discuss the results on the two environmental tasks, tree cover and elevation. Intuitively, we do not expect S2Vec to have strong relative performance here. Compared to the earlier socioeconomic tasks, for such environmental ones satellite imagery provides a much stronger signal relative to infrastructure category features. Our numerical results support this intuition; image-based approaches outperform S2Vec on both tasks. Adding the location signal mildly boosts S2Vec performance, but not enough to compete with the image-based embeddings. Finally, multimodal combinations of S2Vec with imagery does significantly improve on the former but not on the latter, further showing the limits of our approach for such tasks.

*5.4.2 Zero-Shot Geographic Adaptation.* Table 3 reports the $R^2$ metric for all approaches on the zero-shot adaptation task, where the held-out test set is from the eastern part of the US and entirely removed from the training and validation sets. *Here, we find that S2Vec significantly outperforms both SATCLIP and GEOCLIP individually on both datasets.* Between RS-MaMMUT and S2Vec, the former is the best individual mode on the population dataset and the latter is the best individual mode on the median income dataset.

The effect of multimodal fusion in this task type is more pronounced than in the random split task. On the population dataset, the multimodal embedding improves on the image-based embedding in all cases, but only improves on S2Vec in the S2-RS case; S2-RS is the best overall. On the median income dataset, where S2Vec is significantly better than all other approaches, only S2-GEOCLIP (the overall winner) improves on it. Again, for all image-based embeddings in this dataset, fusing with S2Vec improves against the image-only metric.

Table 3. Zero-shot Geographic Adaptation performance on the two US-wide datasets: population and median income. Over 20 independently initialized training runs, we report the mean and standard deviation of the $R^2$ metric (higher better) over the held-out test set.

| Method | Population $R^2 \uparrow$ | Median Income $R^2 \uparrow$ |
|---|---|---|
| *Unimodal* | | |
| S2Vec | 0.64 ± 0.004 | 0.45 ± 0.006 |
| SATCLIP | -0.6 ± 0.35 | -7.24 ± 0.87 |
| GEOCLIP | 0.33 ± 0.01 | 0.30 ± 0.02 |
| RS-MaMMUT | 0.68 ± 0.008 | 0.23 ± 0.03 |
| *Multimodal* | | |
| S2-SATC proj-add | 0.56 ± 0.01 | 0.37 ± 0.01 |
| S2-GEOC proj-add | 0.6 ± 0.01 | **0.48** ± **0.01** |
| S2-RS proj-add | **0.72** ± **0.006** | 0.35 ± 0.007 |

## 6 Discussion

Our work highlights several key insights regarding geospatial representation learning. First, S2Vec works much better in the geographic adaptation task than in the random split task. Zero-shot geographic adaptation, where the model is tested on a completely unseen region, is in some ways more difficult and arguably more important for making models that can work anywhere in the world – a key goal for geospatial foundation models. The strong performance of S2Vec here shows the value of encoding built environment features in a way that captures spatial relationships and depends less on specific regional patterns learned from training data.

Second, the results from our multimodal fusion experiments present a mixed but useful signal. Usually, combining S2Vec with image data improves the results compared to using either one alone, suggesting that they effectively complement each other. However, this improvement doesn't happen when one of the individual methods is already very good or the best at a task. Also, note that in our experiments, the embeddings are fused after independent pre-training. Such downstream fusion may limit how much they help each other compared to combining them during pre-training.

Third, S2Vec uses a lightweight approach by leveraging built environment feature vectors and masked autoencoding. Despite being relatively simple in comparison to the image-based models, S2Vec achieves competitive performance across various socioeconomic prediction tasks. This effectiveness is at least in part due to the hierarchical multi-resolution aspect of the S2 Geometry Library, which allows for efficient partitioning of large geographic areas into cells of varying resolutions. This hierarchical structure enables S2Vec to capture both fine-grained local features and broader spatial relationships by rasterizing feature vectors from fine-grained cells into images of coarser-grained cells.

Overall, S2Vec and our results are important for how we learn to represent geospatial data and build foundation models. S2Vec offers a scalable and versatile approach to encode built environment information, yielding strong performance across various socioeconomic prediction tasks. We also show how it is useful to combine different types of data and different encoding techniques to capture the complexity of geospatial data. As we work towards more general and powerful geospatial foundation models, methods like S2Vec give us valuable tools and ideas for representing and understanding the world around us.

**Scope and Limitations** The current implementation of S2Vec uses count-based feature vectors of PoI categories and infrastructure elements. While our results demonstrate that this approach is effective for capturing the functional mix and density of the built environment, there are inherent limitations. S2Vec would fall short in applications that need more specialized data. Some of our experiments demonstrated these shortcomings on environmental tasks such as tree cover and elevation. Other tasks related to mobility, such as assessing walkability or traffic properties of streets or neighborhoods, and broader segmentation and classification tasks would need richer and more diverse feature vectors that we propose as future work. The raw counts also do not capture the quality, scale, or relevance of geospatial entities. For example, a large and popular commercial center is currently treated the same as a small strip of rarely-visited shops, and the topology of the road network is simplified into a single count. This means that nuanced geospatial insights related to the economic significance of a PoI or the geometric layout of a neighborhood are currently beyond the scope of our features.

## 7 Conclusion

In this paper, we introduced S2Vec, a scalable self-supervised framework for learning geospatial embeddings of the built environment. S2Vec uses the S2 Geometry Library for efficient spatial partitioning, rasterizes built environment feature vectors, and learns powerful and task-agnostic representations with masked autoencoding. Our evaluations on downstream socioeconomic prediction tasks demonstrate that S2Vec is competitive with image-based embedding methods and that multimodal fusion with image-based embeddings often leads to improved performance, highlighting the complementary nature of different geospatial data modalities.

Our work lays a foundation for several promising avenues of future research. First, we could use significantly richer feature vectors to represent a cell and location. Beyond counts of built environment features, future work could explore encoding other valuable geospatial information, such as overhead building geometries, elevation profiles, and time-dependent features related to mobility patterns and traffic flow. Having such diverse data sources would lead to even more comprehensive and informative embeddings.

Second, while this work uses the masked autoencoding approach inspired by the Vision Transformer, other architectures could be explored that better leverage structured geospatial relationships. For example, a Swin transformer [41] can learn more robust representations with MAE. Alternatively, a graph transformer network [43] could explicitly model the complex relationships and dependencies between neighboring cells or other geospatial entities.

Third, while this work uses a fixed parent-child resolution, future work could systematically explore different combinations of S2 levels. This would help in understanding the optimal trade-off between fine-grained local detail and broad spatial context for different tasks and geographic regions. An even more exciting direction would be to develop a multi-scale masked autoencoding approach that explicitly leverages the full S2 hierarchy, allowing the model to learn from several resolutions simultaneously for more robust and comprehensive geospatial embeddings.

Fourth, the core S2Vec framework could be adapted to learn spatially aware embeddings of other geospatial entities beyond grid cells, such as points of interest, road networks, and other vector-based geographic data. Developing methods to effectively represent and relate these different types of geospatial information is a critical challenge in GeoAI.

Finally, an important direction for future work could be joint pre-training strategies. Instead of fusing embeddings from independently pre-trained models, future research could explore joint pre-training with overhead imagery and feature vectors through multi-channel masked autoencoding. This would allow the model to learn cross-modal representations in a more integrated manner, potentially leading to even more powerful embeddings.

## Acknowledgments

The authors would like to thank Pranjal Awasthi, Nadav Sherman, Genady Beryozkin, and Tomer Shekel (all from Google Research) for valuable comments and insights.

## A Point-of-Interest Category Distributions

The S2Vec feature vector uses Google Maps point of interest (POI) categories, similar to the OSM category tags (though they do not always have a clear one-to-one mapping). Figure 5 demonstrates how the counts per s2 cell for different POI categories vary across urban and suburban areas. For instance, in the San Francisco Bay Area, gas stations are concentrated along high-traffic routes. In contrast, they are located more on the periphery of Manhattan, likely due to the city's robust public transportation system. Additionally, gyms and fitness centers are more widely distributed throughout the Bay Area, whereas in Manhattan, they are concentrated primarily in the downtown core.

## B An Example of Clustering Embeddings

To demonstrate how the s2vec embeddings capture a signature of the built environment, we performed k-means clustering (with k=8), again on the San Francisco Bay Area and Manhattan metropolitan areas. The resulting clusters, as visualized in Figure 6, effectively segment the regions. This clustering reveals that areas with similar urban characteristics are grouped together, and they are clearly separated from less populated or uninhabited regions. While a simple clustering cannot fully convey the high-dimensional information captured by the embeddings, it provides a compelling illustration of their ability to capture semantic similarity within a geographic context.

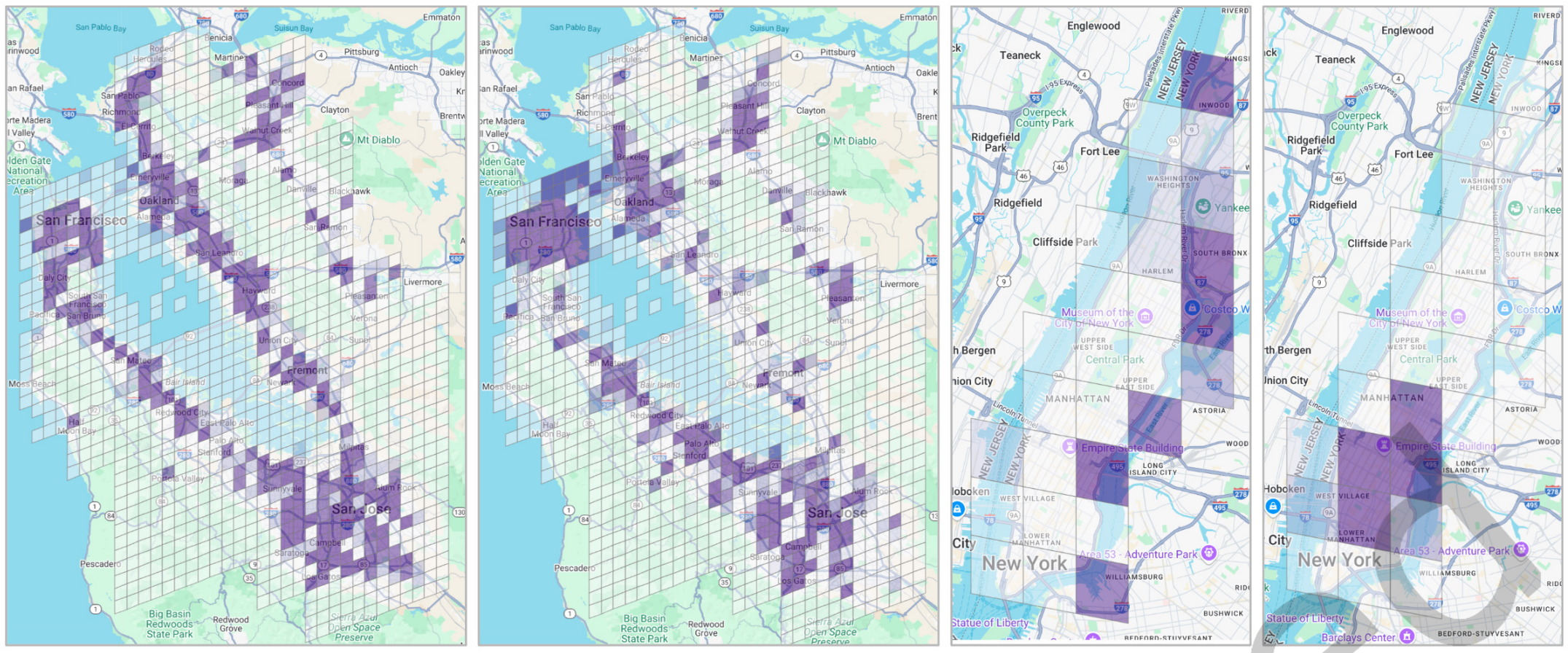

Fig. 5. We illustrate how specific features vary across two large metropolitan urban areas: San Francisco Bay Area (the two left images) and Manhattan, New York City (the two right images). The shading opacity corresponds to the relative proportion of the features (specifically the z-score normalized counts); the first and third images show the distribution of gas stations, and the second and fourth images illustrate gyms and fitness centers.

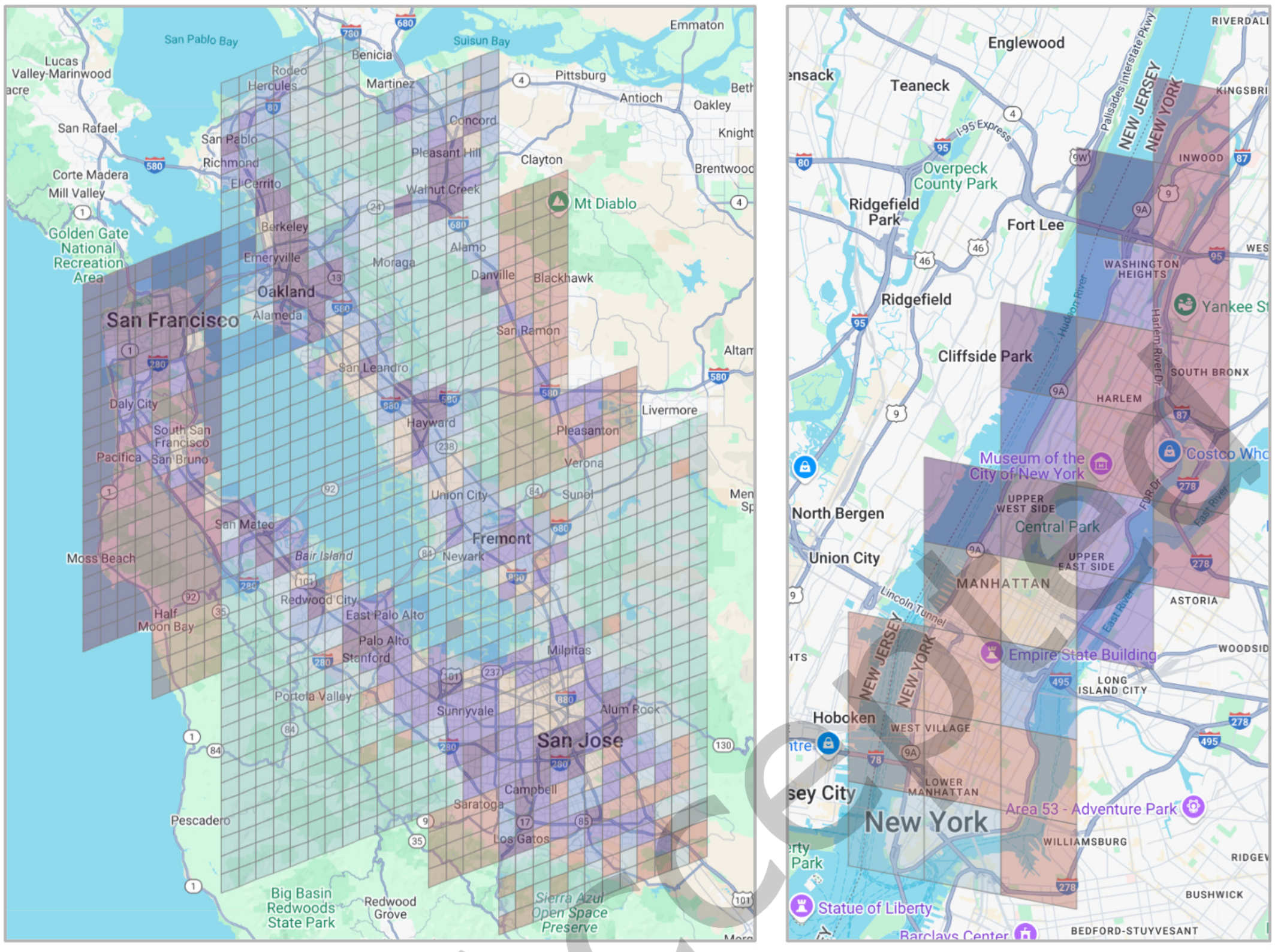

Fig. 6. k-means clustering results for two distinct urban areas, displayed side-by-side. The image on the left shows the clustering for the San Francisco Bay Area, and the image on the right shows the clustering for Manhattan. The clusters are independently derived from S2Vec embeddings for each location and are represented by different colors.